\documentclass[%
 aps,
 amsmath,amssymb,
reprint,%
superscriptaddress
]{revtex4-1}

\usepackage{graphicx}
\usepackage{dcolumn}
\usepackage{bm}

\newcommand{\singletS}{$^1$S$_0\,$}
\newcommand{\tripletP}{$^3$P$_1\,$}
\newcommand{\singletP}{$^1$P$_1\,$}
\newcommand{\tripletPzero}{$^3$P$_0\,$}
\newcommand{\tripletPtwo}{$^3$P$_2\,$}

\begin{document}

\preprint{AIP/123-QED}

\title{Narrow-line cooling and imaging of Ytterbium atoms in an optical tweezer array}

\author{S. Saskin}
    \thanks{These authors contributed equally to this work.}
\affiliation{ 
Department of Electrical Engineering, Princeton University, Princeton, NJ 08540
}%
\affiliation{ 
Department of Physics, Princeton University, Princeton, NJ 08540
}%
\author{J.T. Wilson}
    \thanks{These authors contributed equally to this work.}
\affiliation{ 
Department of Electrical Engineering, Princeton University, Princeton, NJ 08540
}%
\thanks{These authors contributed equally to this work.}
\author{B. Grinkemeyer}
\affiliation{ 
Department of Electrical Engineering, Princeton University, Princeton, NJ 08540
}%
\author{J.D. Thompson}
 \email{jdthompson@princeton.edu}
\affiliation{ 
Department of Electrical Engineering, Princeton University, Princeton, NJ 08540
}%

\date{\today}

\begin{abstract}
Engineering controllable, strongly interacting many-body quantum systems is at the frontier of quantum simulation and quantum information processing. Arrays of laser-cooled neutral atoms in optical tweezers have emerged as a promising platform, because of their flexibility and the potential for strong interactions via Rydberg states. Existing neutral atom array experiments utilize alkali atoms, but alkaline-earth atoms offer many advantages in terms of coherence and control, and also open the door to new applications in precision measurement and timekeeping. In this work, we present a technique to trap individual alkaline-earth-like Ytterbium (Yb) atoms in optical tweezer arrays. The narrow \singletS-\tripletP intercombination line is used for both cooling and imaging in a magic-wavelength optical tweezer at 532 nm. The low Doppler temperature allows for imaging near the saturation intensity, resulting in a very high atom detection fidelity. We demonstrate the imaging fidelity concretely by observing rare ($<$ 1 in $10^4$ images) spontaneous quantum jumps into and out of a metastable state. We also demonstrate stochastic loading of atoms into a two-dimensional, 144-site tweezer array. This platform will enable advances in quantum information processing, quantum simulation and precision measurement. The demonstrated narrow-line Doppler imaging may also be applied in tweezer arrays or quantum gas microscopes using other atoms with similar transitions, such as Erbium and Dysprosium.
\end{abstract}

\pacs{Valid PACS appear here}
\maketitle

Neutral atom arrays are an emerging platform for quantum simulation and quantum information processing. The use of individual optical tweezers \cite{Schlosser:2001vy} to trap atoms offers unprecedented control for “bottom-up” assembly of large-scale quantum systems, while interactions and entanglement can be realized through collisions \cite{Kaufman:2015il}, Rydberg states \cite{Jaksch:2000eg,Isenhower:2010uq,Wilk:2010db,Kaufman:2015il,Jau:2015jn,Lienhard:2018fm,Bernien:2017bp}, optical cavities \cite{Welte:2018be} or the formation of molecules \cite{Liu:2018bn}. Crucially, the entropy associated with stochastic loading from a magneto-optical trap can be eliminated using rapid imaging, feedback and rearrangement of the atoms' positions \cite{Weiss:2004fk}, allowing for uniform filling of large 1D \cite{Endres:2016fk}, 2D \cite{Kim:2016hx,Barredo:2016ea} and 3D \cite{Lee:2016ff,Barredo:2018ee} arrays. In recent years, these systems have been used to probe many-body quantum dynamics \cite{Lienhard:2018fm,Bernien:2017bp} engineer multi-qubit gates, and prepare entangled states \cite{Isenhower:2010uq,Wilk:2010db,Kaufman:2015il,Jau:2015jn}.

\begin{figure}[!b]
\includegraphics{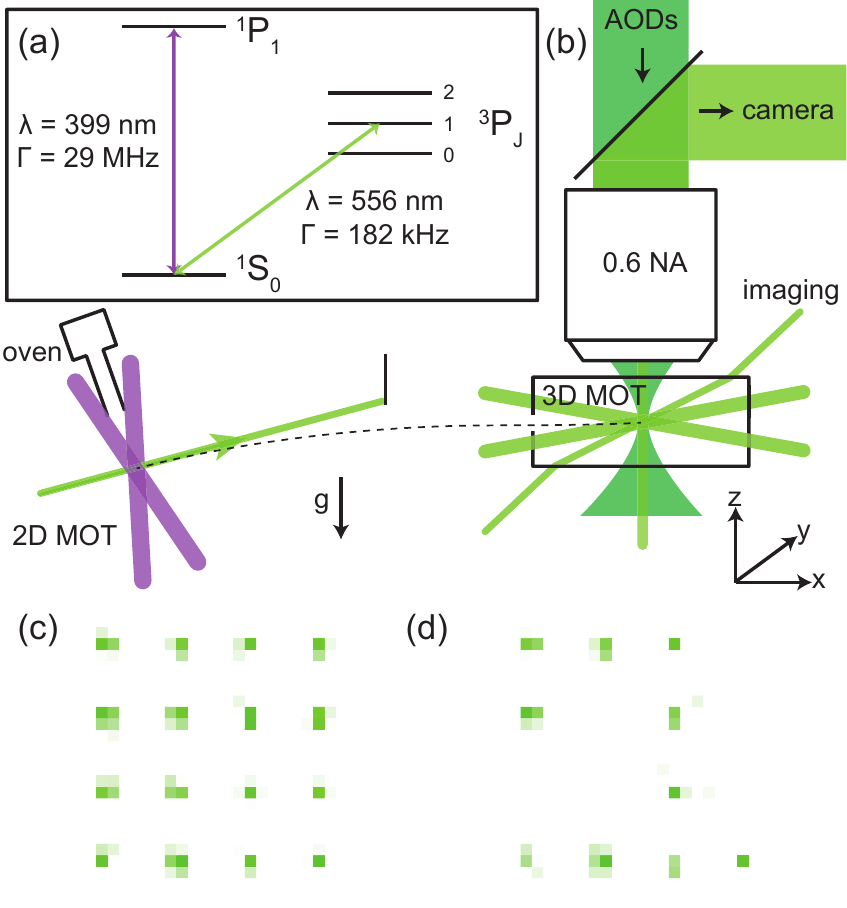}
\caption{\label{fig:schematic}(a) Relevant energy levels for $^{174}$Yb, with transition wavelengths ($\lambda$) and linewidths ($\Gamma$) indicated. (b) Diagram of experimental setup indicating the geometry of the cooling, imaging and trapping beams. Two of the 3D MOT beams are in the $xy$-plane, while the third propagates through the objective lens along the $z$-axis. The angled imaging beam is in the $xz$-plane. For other details, see text. (c) Average and (d) single-shot images of atoms in a 4x4 tweezer array.}
\end{figure}

All experiments to date involving optical tweezers have utilized alkali atoms, in particular Rb \cite{Schlosser:2001vy,Isenhower:2010uq,Kaufman:2015il,Endres:2016fk,Kim:2016hx,Barredo:2016ea,Barredo:2018ee}, Cs \cite{Jau:2015jn,Hutzler:2017kl,Liu:2018bn} and Na \cite{Hutzler:2017kl,Liu:2018bn}.  However, alkaline earth atoms offer several intriguing advantages \cite{Daley:2008ih} including ultra-long coherence for nuclear spins in the $J=0$ electronic ground state, a combination of strong and narrow optical transitions for rapid laser cooling to very low temperatures, and metastable shelving states to facilitate high-fidelity qubit readout. Interaction between nuclear spin qubits can be realized using Rydberg states (which feature strong hyperfine coupling in alkaline earth atoms \cite{Liao:1980fs,Beigang:1983ch}) or coherent spin-exchange collisions using the metastable clock state \cite{Pagano:2018vx,Scazza:2014jw,Zhang:2014el,Cappellini:2014fv}. Furthermore, Rydberg states may be trapped using the polarizability of the alkali-like ion core \cite{Topcu:2013dw}. Lastly, alkaline earth atoms are widely used in optical lattice clocks for precision timekeeping and measurement because of their long-lived metastable states  \cite{Ludlow:2015ul}.

In this work, we demonstrate an approach to produce large-scale arrays of individual alkaline-earth-like Yb atoms trapped in optical tweezers. Both cooling and imaging are performed on the narrow \singletS-\tripletP intercombination line ($\lambda=556$ nm, linewidth $\Gamma = 182$ kHz), enabled by the convenient ``magic'' trapping condition for these states with 532 nm trapping light \cite{Yamamoto:2016cy}. The use of a narrow transition allows rapid cooling to temperatures of 6.4(5) $\mu$K, near the theoretical Doppler temperature of 4.4 $\mu$K for this transition. In contrast to most previous single-atom detection schemes relying on polarization gradient \cite{Schlosser:2001vy}, Raman sideband \cite{Omran:2015io,Parsons:2015ck,Cheuk:2015jr} or EIT \cite{Edge:2015gr,Haller:2015hi} cooling during imaging, the narrow linewidth enables high fidelity imaging in shallow traps using Doppler cooling alone. As an outlook, we demonstrate a 144-site (12x12) tweezer array, stochastically loaded with atoms.

Our experimental apparatus is depicted schematically in Fig. 1b. At the center is a glass cell with primary windows of 2 in. diameter and 6.35 mm thickness. An objective lens designed to compensate for the window thickness, with numerical aperture NA=0.6 (Special Optics), is used to focus the tweezer array and image fluorescence from the trapped atoms. $^{174}$Yb atoms from a needle-collimated oven \cite{Senaratne:2015cu} at 440$^\circ$C are initially cooled in a 2D magneto-optical trap (MOT) operating on the broad 399 nm transition, then accelerated through a differential pumping tube into the glass cell using a push beam on the 556 nm intercombination line \cite{Tiecke:2009hy}. The 2D MOT is connected to the glass cell at an angle, such that the atoms sag 25 mm under gravity during flight, allowing optical line-of-sight between the 2D MOT and the glass cell to be blocked by a pick-off mirror. In the glass cell, the atoms are directly loaded into a frequency-broadened 3D MOT operating on the 556 nm transition, then compressed into a single-frequency MOT to load the optical tweezers. The MOT beams are in an orthogonal six-beam configuration, with the vertical beams passing through the objective lens. We typically load 2 $\times$ 10$^{5}$ atoms with a density of 10$^{11}$ cm$^{-3}$ in 200 ms.

\begin{figure}
\includegraphics{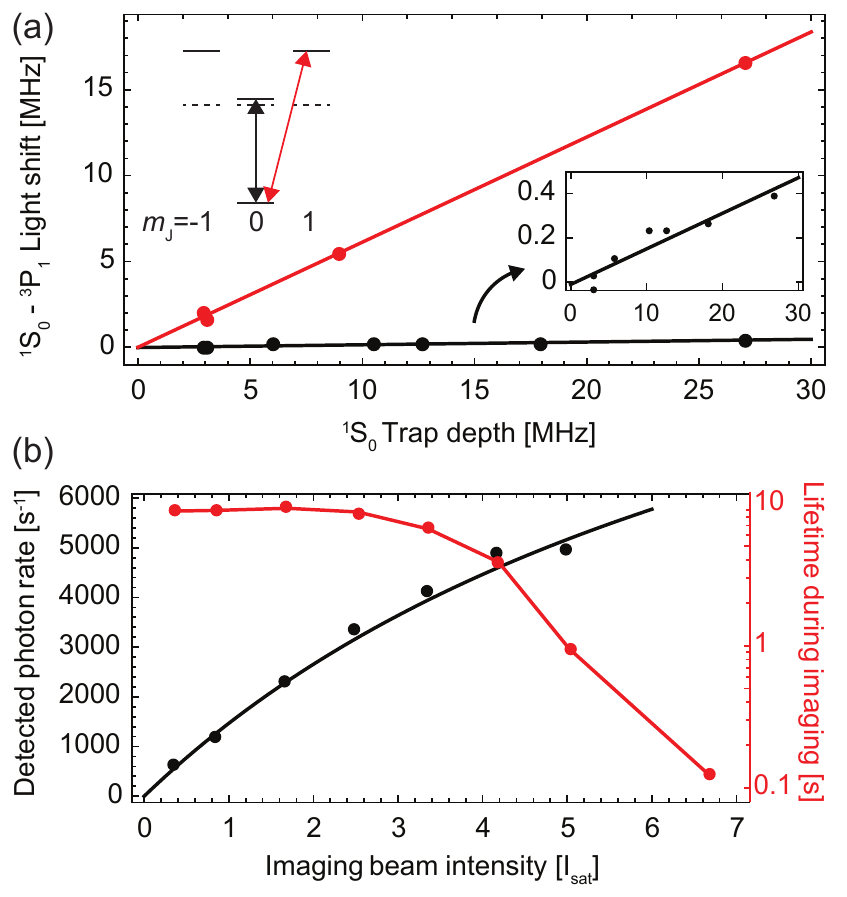}
\caption{\label{fig:temp} (a) Light shifts of the \singletS - \tripletP transition as a function of the optical tweezer power. The tensor light shift lifts the degeneracy of the \tripletP $m_J$ levels, resulting in different light shifts for the $m_J=0$ and $m_J=\pm 1$ excited states. The light shift for the $\Delta m_J=0$ transition is 1.6\% of the ground state trap depth, which corresponds to about $\Gamma/2$ at a typical tweezer depth of 6 MHz (0.29 mK). The horizontal axis is calibrated using the previously measured value of the \tripletP $m_J=\pm 1$ polarizability at 532 nm \cite{Yamamoto:2016cy}. (b) Lifetime and scattering rate of trapped atoms under various imaging intensities at a typical imaging detuning of $\Delta \approx -1.5 \Gamma$. The lifetime decreases exponentially with increasing imaging power above $I/I_{sat} \approx 4$. We find $I/I_{sat} \approx 3$ to be the optimal balance of photon scattering rate and lifetime for this detuning.}
\end{figure}

The optical tweezer array is generated by a pair of orthogonally-oriented acousto-optic deflectors \cite{Kaufman:2015il,Endres:2016fk}, driven by arbitrary waveform generators. The tweezers are focused to a beam waist of approximately 700 nm, with 6 mW of power per tweezer at the input to the objective, yielding a trap depth of 6 MHz (0.29 mK). After overlapping the compressed MOT with the tweezers for a loading time of 30 ms, the 3D MOT beams are turned off and the trapped atoms are imaged using a retro-reflected beam propagating diagonally with respect to the tweezer propagation direction, with projection onto both the radial and axial oscillation directions. Fluorescence from the atoms is collected through the objective and imaged onto a sCMOS camera (Photometrics Prime BSI). Average and single-shot images of a 16-site (4x4) array are shown in Fig. 1c,d.

\begin{figure*}[t]
\includegraphics[width=178mm]{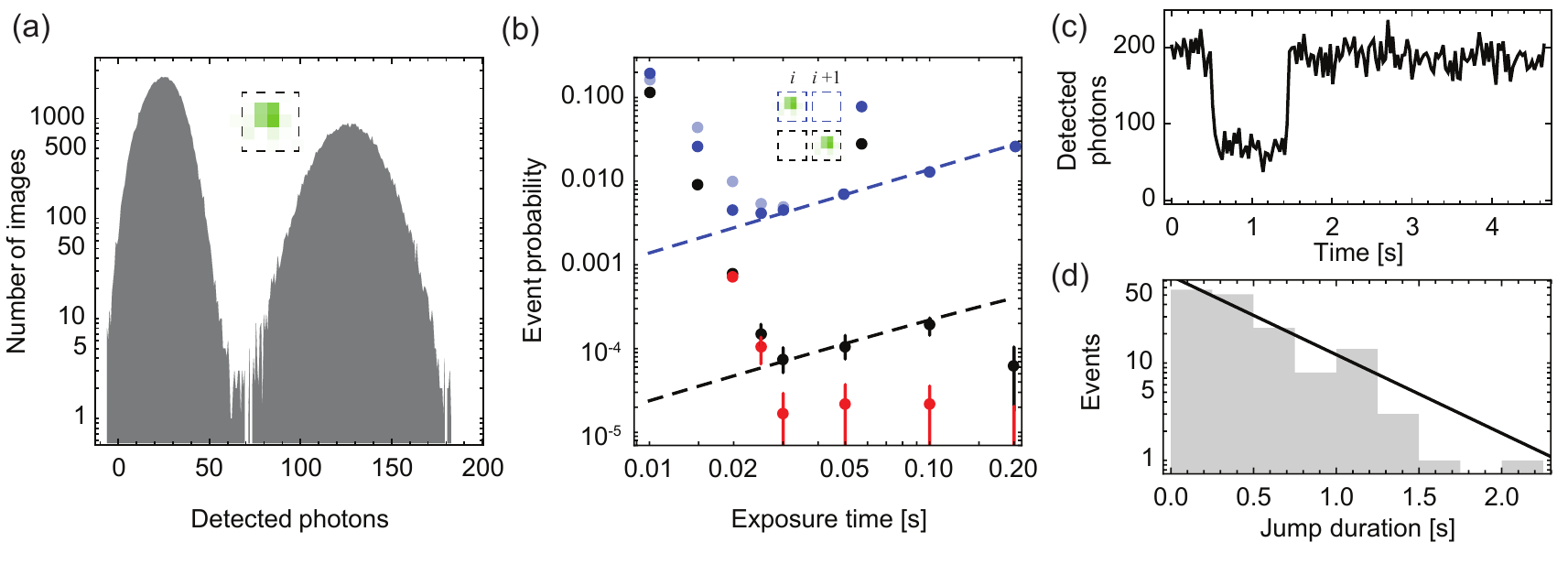}
\caption{\label{fig:array} (a) Histogram of detected photons at a given site for an exposure time of 30 ms ($\sim$ 136,000 images), revealing clear separation between fluorescence counts for 0 and 1 atoms per site. (b) Imaging fidelity, quantified by the probability of disagreement between two subsequent images of the same array. Two event types are classified: blue points show the probability of bright sites appearing dark in the next image [$P_{b\rightarrow d}=P(n_{i+1}=d|n_i=b)$, where $n_i$ denotes the state in image $i$] and black points show the probability of a dark site appearing bright in the next image [$P_{d\rightarrow b}=P(n_{i+1}=b|n_i=d)$]. The light blue symbols show the classification using a simple count threshold, while the other points (blue, black, red) use a pixel-wise Bayesian classifier that has approximately half the error rate. For exposure times greater than 20 ms, $P_{b\rightarrow d}$ is dominated by atom loss, consistent with the independently measured lifetime (7.2 s) for these imaging conditions (blue curve). $P_{d\rightarrow b}$ reaches a floor below $1 \times 10^{-4}$ that originates from quantum jumps out of a metastable state. A representative jump event is shown in panel (c): a tweezer initially loaded with an atom goes dark, but spontaneously becomes bright one second later, though the MOT is off the entire time. The duration of these events [panel (d)] is consistent with a metastable state lifetime of $\tau_m = 0.54(7)$ s. The black dashed curve in (b) is a fit to $P_m(1-e^{-t/\tau_m})$, which describes the rate of these events for an average metastable state population $P_m$, which we infer to be $P_m = 4 \times 10^{-3}$. The red points in (b) show $P_{d\rightarrow b}$ with conclusively identified quantum jump events removed; no more than two errors remain in roughly 180,000 images at each exposure time 30 ms or longer.}
\end{figure*}

To characterize the cooling and imaging properties of the 556 nm transition, we first measure the differential light shift of the \singletS and \tripletP states in the optical tweezers (Fig. 2a). In the absence of a magnetic field and with linearly polarized trapping light, the tensor light shift lifts the degeneracy of the \tripletP $m_J$ states, resulting in different potentials for $m_J=0$ and $m_J= \pm1$, quantized along the light polarization direction. We measure the frequencies of the $m_J=0$ and $m_J= \pm 1$ transitions as a function of trap depth by blowing atoms out of the trap using resonant light, and observe that the $m_J=0$ transition shifts by approximately 1.6\% of the ground state trap depth, in agreement with previous measurements \cite{Yamamoto:2016cy}. Under typical trapping conditions, the transition frequency is blue-shifted $90$ kHz $\approx \Gamma/2$ in the trap. The positive sign and small magnitude of this shift facilitates efficient loading of atoms from the \tripletP MOT into the tweezers.

After loading the tweezers and applying a brief pulse to remove multiple atoms (20 ms, $\Delta \approx -2\Gamma$, $I/I_{sat}\approx5$), we measure an atomic temperature of $6.4(5)\,\mu$K (using the release-and-recapture technique \cite{Reymond:2003fo}). In order to determine the optimal fluorescence imaging parameters, we study the lifetime of the trapped atoms in a 0.29 mK deep potential under continuous illumination from the imaging beam as a function of intensity at a detuning $\Delta = -1.5 \Gamma$ (Fig. 2b). The lifetime decreases exponentially with intensity, consistent with a linear increase in temperature \cite{Lett:1989vy} and exponentially-activated tunneling over a barrier; however, at moderate intensities ($I/I_{sat} \approx 3$) we achieve lifetimes near 10 seconds with a photon scattering rate that we estimate to be $0.29 \times \Gamma/2$. The measured temperature during imaging is 13(2) $\mu$K. In deeper traps, we observe longer lifetimes at high imaging intensities, consistent with the model of heating-induced loss. A histogram of the number of detected photons on a single site during a 30 ms exposure is shown in Fig. 3a.

An important metric for initializing large-scale low-entropy arrays and performing high-fidelity qubit readout is the fidelity with which a single atom can be imaged. To quantify this, we take repeated images of a 9-site (3x3) array for 5 seconds under continuous illumination, with varying exposure time and negligible delay between images. In each image, we classify each site to be either bright or dark, indicating the presence or absence of an atom; ideally, this would remain unchanged across multiple images. We quantify the imaging performance by the probability of either of two events to occur: $P_{b\rightarrow d} = P(n_{i+1}=d|n_i=b)$, indicating that a bright site transitions to dark in the next image, and $P_{d\rightarrow b} = P(n_{i+1}=b|n_i=d)$, indicating that a dark site appears bright in the next image.

At short exposure times, both events occur often because of noise. At exposure times greater than 20 ms, $P_{b\rightarrow d}$ is limited by loss from the traps, in a manner consistent with the independently measured lifetime of 7.2 s for these imaging conditions. The minimum value ($P_{b\rightarrow d} = 4.5(3) \times 10^{-3}$, averaged across all sites in the 3x3 array) occurs at 20 ms imaging time.

Interestingly, for exposure times greater than 25 ms, nearly all of the $d\rightarrow b$ events are followed by multiple bright images, suggesting that the transition was not the result of statistical uncertainty (\emph{i.e.}, misclassification of a bright image), but by the sudden appearance of an atom. We hypothesize that these events correspond to quantum jumps of atoms from metastable states back to the ground state. The measurement record of one such event is shown in Fig. 3c, showing a bright site transitioning to dark and back to bright. A histogram of the duration of many such events (158 events captured over approximately $5 \times 10^{4}$ atom-loading events, Fig. 3d) reveals the dark state lifetime to be $\tau_m = 0.54(7)$ s. This is much shorter than the natural lifetime of the metastable \tripletPzero or \tripletPtwo states ($>$ 10 s), but is roughly consistent with an estimate of the rate of off-resonant scattering of dipole trap photons from these states, which would return the atoms to \tripletP and eventually \singletS (a similar mechanism is presumably responsible for populating the metastable state in the first place). An alternative interpretation, loading of new atoms from the background vapor, is ruled out by the fact that these events are nearly always preceded by a $b \rightarrow d$ transition. From the rate of these events, we infer that the average fraction of atoms in the involved metastable states is $P_m = 4 \times 10^{-3}$. Determining which state(s) are involved and how they are populated will be the subject of future work.

These quantum jump events are readily classified by the appearance of more than one bright image following more than one dark image. By excluding these events (which are not imaging errors \emph{per se}) from $P_{d\rightarrow b}$, we compute a new error rate describing statistical uncertainty of the image assignment, shown in red in Fig. 3b. Our dataset of $1.8\times10^5$ images with 30 ms exposure time contains only two such events; since atoms are present in 30\% of all images in this dataset, this is equivalent to a statistical atom detection uncertainty of $1 - 3(3) \times 10^{-5}$. This extremely high fidelity is crucial for observing the quantum jump events shown in Fig. 3c, which occur in fewer than 1 in $10^4$ images at this exposure time.

The quantity $P_{b\rightarrow d}$ is important because it sets an upper bound on the size of the atom array that can be filled without defects ($N_{max} \approx 1/P_{b\rightarrow d}$), since atoms must survive the initial image (additional contributions arise from the rearrangement process itself \cite{Endres:2016fk,Barredo:2016ea}). Our value, $P_{b\rightarrow d} = 4.5(3) \times 10^{-3}$ ($N_{max} \approx 220$), is comparable to the lowest directly measured quantity reported in the literature, despite our use of a narrow transition for imaging (previously, values around 0.006-0.01 have been reported \cite{Endres:2016fk,Picken:2017dd}). Two factors may contribute to this surprising result. First, the intrinsic low Doppler temperature allows imaging close to saturation, so the count rates we observe are only a factor of 2-4 lower than those obtained with polarization gradient cooling in Rb in similar traps \cite{Endres:2016fk}. Second, our sCMOS camera is nearly shot-noise limited even for small photon numbers, and should theoretically have lower noise than an EMCCD when the photon number per pixel is greater than $\approx 5$ \cite{Picken:2017dd}.


\begin{figure}[t]
\includegraphics{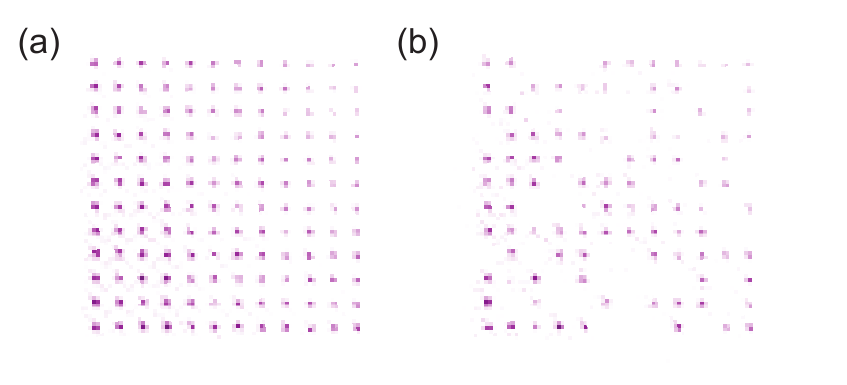}
\caption{\label{fig:temp} (a) Average and (b) single-shot images of a 12x12 tweezer array using simultaneous \singletP imaging and \tripletP cooling. The detected photon rate is much lower for this imaging method, so the exposure time is 500 ms. Over repeated single-shot images, the average (worst) site has loading probability $p=0.49$ ($p=0.35$).}
\end{figure}

As an outlook, we demonstrate stochastic loading of a 144-site (12x12) array of optical tweezers (Fig. 4). Auto-fluorescence from the trap light (proportional to the total number of tweezers) prevents us from imaging this array at 556 nm using the techniques described above. However, there is very little trap-induced fluorescence at 399 nm (higher in energy than 532 nm), which enables us to image scattered light from the \singletP transition while simultaneously cooling on the \tripletP transition, following Ref. \cite{Yamamoto:2016cy}. Modifying the optical setup to reduce the overlap of the trapping and imaging paths, and improving spatial and spectral filtering, will enable imaging large-scale arrays with 556 nm light.

The narrow-line imaging demonstrated here indicates that narrow lines with $\Gamma \approx 200$ kHz are a ``sweet spot" for single-atom fluorescence imaging in optical traps, offering a balance between photon detection rate and low temperatures during imaging. This may be applied to optical tweezer arrays and quantum gas microscopes based on other atomic species with similar transitions, including Er \cite{Frisch:2012bb} and Dy \cite{Lu:2011gc}.

Ytterbium optical tweezer arrays create several new opportunities for quantum simulation and quantum computing. In particular, $^{171}$Yb is a promising qubit as its $I=1/2$ nuclear spin should have exceptional coherence, with low sensitivity to magnetic field noise and differential light shifts in deep optical traps. Furthermore, $m_I$-selective shelving in the metastable $^3$P$_0$ or $^3$P$_2$ states will allow the demonstrated high-fidelity atom detection to be translated into high-fidelity state detection. The $^3$P$_0$ state may also be used as a starting point for single-photon excitation to the Rydberg states ($\lambda=302$ nm). Interestingly, storing quantum states in $^3$P$_0$ may also allow site-selective non-destructive measurement by transferring individual atoms to \singletS: repeated driving on the \singletP and \tripletP transitions will not perturb the nuclear spin states of atoms remaining in $^3$P$_0$. Similarly, these states would be protected from fluorescence from a nearby MOT used to continuously replace lost atoms.

The level structure of the $^{171}$Yb Rydberg states also offers several interesting properties for quantum information and simulation. Strong hyperfine coupling emerges between the nuclear spin and the Rydberg electron in the $6snl$ Rydberg states, mediated by the hyperfine coupling to the core electron and the singlet-triplet splitting energy \cite{Liao:1980fs}. This coupling can be utilized to directly realize two-qubit entanglement and gates involving the nuclear spin, and will also create new possibilities for implementing interacting spin models using Rydberg dressing \cite{Glaetzle:2015iu}. A complete characterization of the Yb Rydberg series is the subject of ongoing work \cite{Lehec:2018tl}. We have recently observed the $6sns$ $^3$S$_1$ series for the first time using MOT depletion spectroscopy, via two-photon excitation through the \tripletP state.

Lastly, tweezer arrays may prove beneficial for improving the performance of neutral Yb optical lattice clocks \cite{Beloy:2014bn}, for example by generating squeezed states using Rydberg interactions, or maintaining multiple sub-ensembles to reduce the impact of local oscillator noise \cite{Rosenband:2013vp}.

We gratefully acknowledge helpful conversations with Manuel Endres, Shimon Kolkowitz and Trey Porto. This work was supported by the Army Research Office (contract W911NF-18-1-0215). J.W. is supported by an NSF GRFP.

\emph{Note} While completing this work, we became aware of related publications on Strontium optical tweezer arrays \cite{Cooper:2018wy,Norcia:2018wp}.

\bibliography{ybv4}

\end{document}